\documentclass[a4paper,12pt]{snicpr}  
\usepackage[latin1]{inputenc}  
\usepackage{a4wide}                                
\usepackage{amsmath}
\usepackage{vmargin}                               
\usepackage{graphicx}                              
\usepackage{color}
\usepackage{subfigure}

\newcommand{\grad}{\nabla}

\begin{document}


\article
{Core transport studies in fusion devices}           
{P\"ar Strand\inst{1}\inst{2},                      
 Andreas Skyman\inst{1} and                
 Hans Nordman\inst{1}}                      
{Department of radio and space science, Chalmers university of technology, EURATOM-VR association\\    
 SE-412 96 G\"oteborg, Sweden &                                    
 EFDA Task Force on Integrated Tokamak Modelling}                                     
{elfps@chalmers.se}                             
{http://www.chalmers.se/rss/EN/research/research-groups/transport-theory/}                                




\section{Introduction}
Comprehensive first principles modelling of fusion
plasmas is a numerically challenging: the complicated magnetic geometry
and long range electromagnetic interactions between multiple species
introduce complex collective behaviour in the plasma.  In addition,
steep density and temperature gradients combined with an inhomogeneous 
magnetic field drives instabilities, resulting
in non-linear dynamics and turbulence.

The turbulence in magnetically confined fusion plasmas has important
and non-trivial effects on the quality of the energy confinement.  
These effects are hard to make a quantitative assessment of
analytically.  The problem investigated in this article is the
transport of energy and particles, in particular impurities, in a
Tokamak plasma.  Impurities from the walls of the plasma vessel cause
energy losses if they reach the plasma core.  It is therefore
important to understand the transport mechanisms to prevent
impurity accumulation and minimize losses.  This is an area of
research where turbulence plays a major role and is intimately
associated with the performance of future fusion reactors, such as
ITER.

With the rapid growth and increased accessibility of high
performance computing (HPC) over the last few decades, plasma
modelling has matured towards an increased predictive capability.  
Particular emphasis has been put on simulation of drift wave
physics, widely accepted as the main source of transport in
the plasma core.  Theory, reduced physics as well as first principles
modelling, and software are developed in a coordinated European
effort to produce a virtual Tokamak, a tool that will become
indispensable, both when it comes to developing and running ITER,
and in the planning of future reactors aimed at energy
production~\publication{ParFusEngDes}.

\section{Physical background}
To arrive at a set of equations that are both meaningful and solvable, some
approximation is necessary.  The advances in high performance computing have
allowed fusion modellers to move from fluid descriptions of the plasma to
kinetic descriptions as the basis for turbulence modelling.  In \emph{kinetic
theory} the plasma is described through distribution functions of velocity and
position for the plasma species.  Hence, kinetic equations are
inherently six-dimensional, however, magnetically confined particles
are constrained to tight orbits along field lines.  This motivates averaging
over the gyration, reducing the problem to five-dimensional \emph{gyrokinetic}
equations\reference{Merz2008}

%

\noindent This is a considerable gain, and the foundation of
most current plasma codes.  In this project, GENE, a
European code developed by IPP-Garching\reference{GENE}, has been used.  GENE
employs a second order accurate explicit finite difference scheme, and
has demonstrated excellent parallel performance using in excess of $10000$
cores\reference{Jenko2000}.

\section{Modelling plasmas}
As mentioned above, gradients drive turbulence.  Here, plasma core turbulence
induced by the so called \emph{ion temperature gradient} (ITG)
mode\reference{Weiland2000}, has
been studied.  Parameters were taken from discharge \#67730 of the \emph{Joint 
European Torus} (JET).  A slice of the simulation domain, illustrating the
turbulence, is shown in figure~\ref{fig:cont}.

The GENE code employs a fixed grid in five dimensional phase space
and a flux-tube geometry.  For a typical simulation for main ions and one
trace species, with electrons considered adiabatic, a
resolution of $n_{x} \times n_{ky} \times n_{z} = 48 \times 48
\times 32$ grid points in real space and of $n_{v} \times n_{\mu} =
64 \times 12$ in velocity space is necessary.  A normal run with
these parameters uses a minimum of $8000$ CPU hours and $384$ cores.
Incorporating kinetic electrons increases the demand for high
resolution in all of phase space, but most notably in velocity
space, and also requires a shorter time step.  Typically, such simulations
require $40000$ CPU hours, occupying $1024$ cores.

The computations produce tens of gigabyte of data to be analysed.  The
non-linear data presented in figure~\ref{fig:PF0} is the result of
approximately twenty runs on the HPC cluster \emph{Akka}, and from the
discussion above, it is readily understood that HPC is vital for this kind of
study.

\section{Transport in ITER-like plasmas}
In general, transport of a species with atomic number $Z$ can locally
be described by a diffusive and a convective contribution.  The former is
characterized by the diffusion coefficient $D_Z$, the latter by a convective
velocity or ``pinch'' $V_Z$.  The \emph{zero flux peaking factor}, defined as
$PF_0 = -R\,V_Z/D_Z$, is important in reactor design because it quantifies the
balance of convective and diffusive transport.  This can be understood from
equation \eqref{eq:transport}, where $\Gamma_Z$ is the impurity flux, $n_Z$ the
density of the impurity species and $R$ the major radius of the
tokamak~\publication{HansArtikel}.  For the regime studied $\grad n_Z$ is
regarded as a constant, such that $-\grad n_Z/n_Z=1/L_{n_Z}$. Setting $\Gamma_Z =
0$ in equation~\eqref{eq:transport} yields the interpretation of $PF_0$ as the
gradient at which the impurity flux vanishes.

\vspace{-1\medskipamount}
\begin{equation}
  \label{eq:transport}
  \Gamma_Z = -D_Z\grad n_Z + n_Z V_Z \Leftrightarrow \frac{R\Gamma}{n_Z} =
D_Z\frac{R}{L_{n_Z}} + RV_Z
\end{equation}

\vspace{-\smallskipamount}
A positive sign of $PF_0$ indicates a net inward transport.  This
might lead to an accumulation of wall impurities in the plasma core,
which can seriously hamper the efficiency of the fusion device.  
Understanding under what circumstances a negative peaking factor can be
achieved is therefore an important issue for ITER and future fusion reactors.

Time series were generated by GENE on the \emph{Akka} HPC cluster for multiple
values of $L_{n_Z}$, and from these $\Gamma_Z$ was extracted.  The parameters 
$D_Z$ and $RV_Z$ were estimated and the peaking factor calculated as their
quotient.  This was repeated for several different nuclear charges $Z$.
Results have been reported in \publication{HansArtikel},
\publication{HansEPS} and \publication{SkymanEPS}.

\begin{figure}[ht]
 \centering
 \subfigure[crossection of the Deuterium density profile showing turbulent
 features]{\includegraphics[height = 54mm]{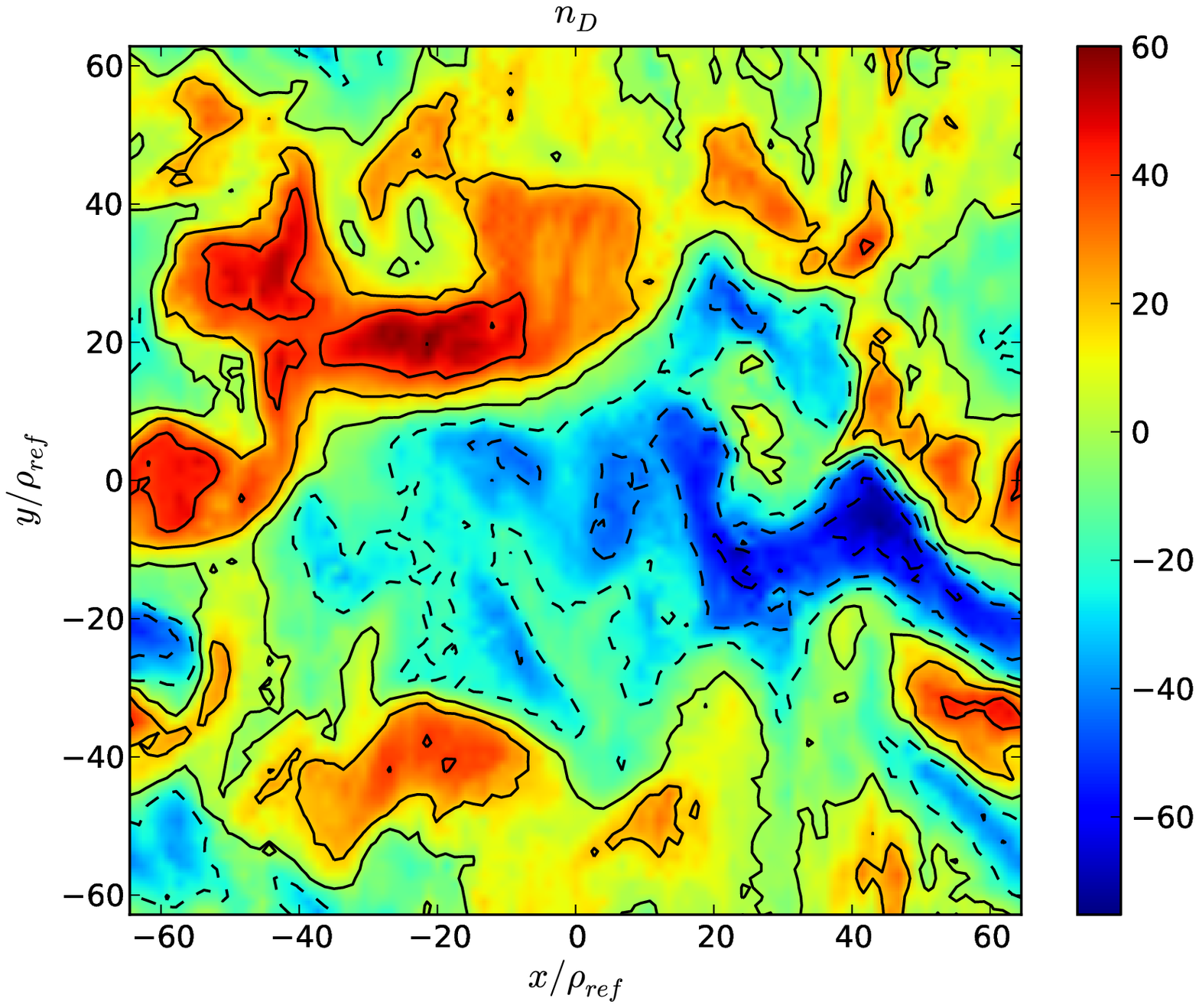} \label{fig:cont}} 
 \subfigure[zero-flux peaking factor for different $Z$ according to three
 different models with adiabatic and one with kinetic
 electrons]{\includegraphics[height = 54mm]{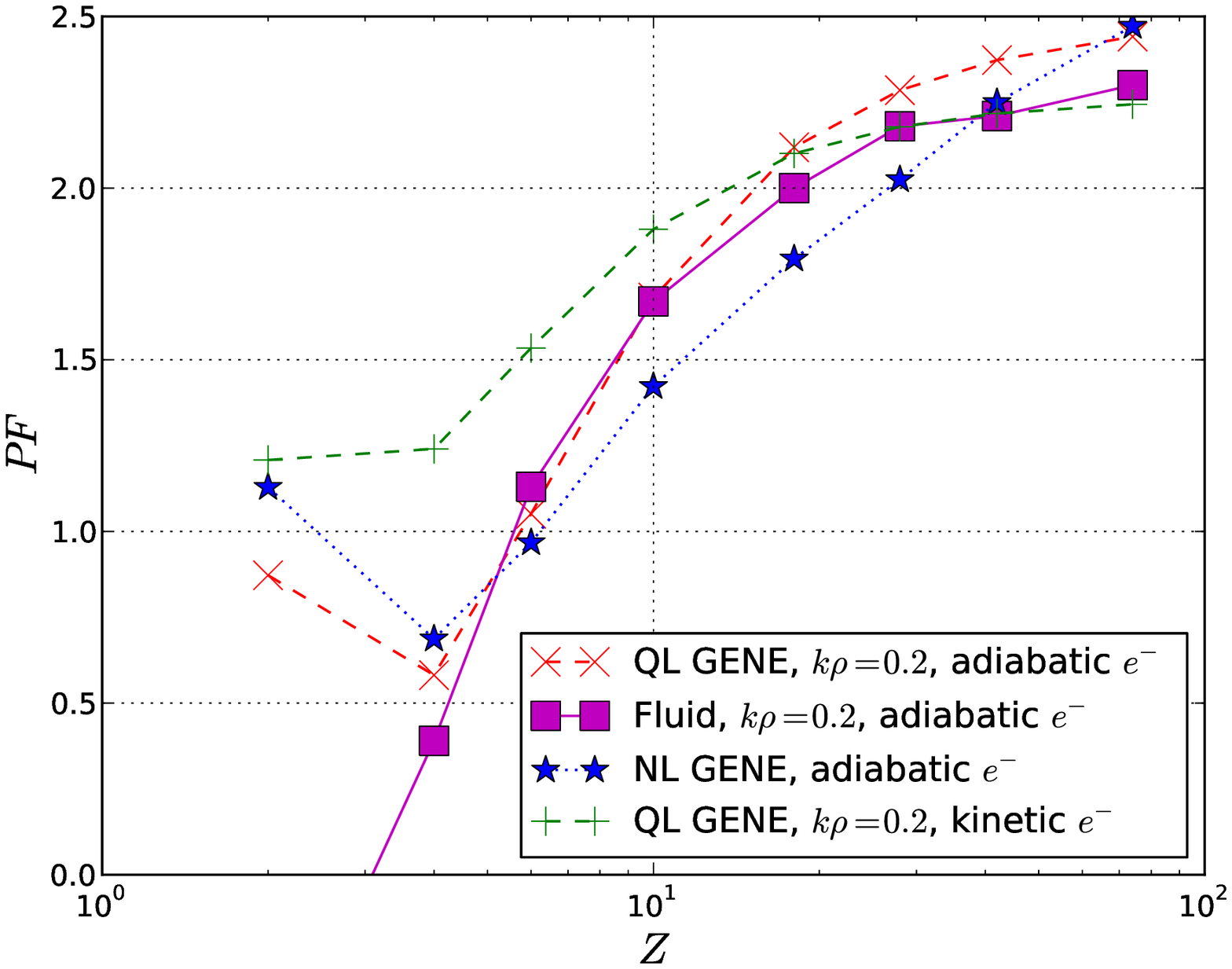} \label{fig:PF0}}
 \caption{Results from the non-linear simulations on Akka}
 \label{fig:thefig}
\end{figure}

Investigating where different models are in agreement is one of the aims of
this kind of study: it is the first step towards an understanding of the
physics that underlie their differences.  In figure~\ref{fig:PF0} the
non-linear results from \emph{Akka} are compared with quasi-linear kinetic
results and results from a fluid model developed at
Chalmers\reference{Weiland2000}.  As can be seen, the three models are in
good qualitative agreement with one another for $Z > 4$ (Be). This owes to the
domination of the single strong ITG mode for these particular parameters, and it
cannot be guaranteed to hold for other cases.  Also in figure~\ref{fig:PF0}, the
quasi-linear result with kinetic electrons is shown for comparison.  From that
one expects kinetic effects to be most pronounced for low $Z$.  At the time of
writing, the fully kinetic non-linear case is still being simulated.

\vspace{-6pt}



\begin{thepublications}{00}
\pubitem{ParFusEngDes} {P.~Strand et al.: 
 \textit{ Simulation \& high performance computing -- building a predictive
 capability for Fusion}, Fusion Engineering and Design (accepted), 2010}
\pubitem{HansArtikel} {H.~Nordman, A.~Skyman, P~Strand et al:
 \textit{Fluid and gyrokinetic simulations of impurity transport in JET},       
 (manuscript), 2010}
\pubitem{HansEPS} {H.~Nordman, A.~Skyman, P.~Strand et al.:
 \textit{Modelling of impurity transport experiments at the Joint European
 Torus}, Proceedings of EPS 2010 (accepted), 2010}
\pubitem{SkymanEPS} {A.~Skyman, H.~Nordman, P.~Strand et al.:
 \textit{Impurity transport in ITG and TE mode dominated turbulence},
 Proceedings of EPS 2010 (accepted), 2010}
\end{thepublications}

\begin{thereferences}{00}
\refitem{Merz2008} {F. Merz:
 \textit{Gyrokinetic Simulation of Multimode Plasma Turbulence}, PhD thesis,
 Westf\"alischen Wilhelms-Universit\"at M\"unster, 2008}
\refitem{Jenko2000} {F. Jenko et al.: 
 \textit{Electron temperature gradient driven turbulence}, Physics of Plasmas,
 \textbf{7}, pp.~1904--10, 2000}
\refitem{Weiland2000} {J. Weiland:
 \textit{Collective Modes in Inhomogeneous Plasma}, Institute of Physics
 Publishing, London (UK), 2000}
\refitem[\ddag]{GENE} {http://www.ipp.mpg.de/\textasciitilde fsj/gene/}
\end{thereferences}



\end{document}